\documentclass[12pt]{article}
\usepackage{float}
\usepackage{subfloat}
\pdfoutput = 1
\usepackage{graphics}
\usepackage{graphicx} 
\usepackage{amsmath}
\usepackage{amsfonts}   
\usepackage{amssymb}
\usepackage{multirow}
\usepackage{hhline}
\usepackage{color}
\usepackage{comment}

\begin{document}
\title{{\bf{\Large   Influence of the cosmological constant on $\kappa$-deformed Neutron Star}}}

\author{ {\bf {\normalsize Bhagya. R}\thanks{22phph03@uohyd.ac.in, r.bhagya1999@gmail.com},\
{\bf {\normalsize Diganta Parai}\thanks{digantaparai007@gmail.com}},\,
{\bf {\normalsize Harsha Sreekumar}\thanks{22phph01@uohyd.ac.in, harshasreekumark@gmail.com}},\,
{\bf {\normalsize Suman Kumar Panja}\thanks{19phph17@uohyd.ac.in, sumanpanja19@gmail.com}}}\\
{\normalsize School of Physics, University of Hyderabad}\\{\normalsize Central University P.O, Hyderabad-500046, Telangana, India}\\[0.2cm]  
}
\date{}

\maketitle
\begin{abstract}
We study a model of the neutron star in $\kappa$-deformed space-time in the presence of the cosmological constant ($\Lambda$). The Einstein tensor and the energy-momentum tensor are generalized to $\kappa$-deformed space-time and we construct the field equations with the cosmological constant. Considering the interior of the star to be a perfect fluid as in the commutative case, we find the Tolman-Oppenheimer-Volkoff equations with the inclusion of the cosmological constant in $\kappa$-deformed space-time. The behavior of the maximum allowed mass of the star and its radius are studied with the variation in the cosmological constant as well as the deformation parameter. We see that the non-commutativity enhances the mass of the star and its maximum mass increases with a decrease in the cosmological constant.
The maximum mass varies from $3.44M_{\odot}$ to $3.68M_{\odot}$ as $\Lambda$ varies from $10^{-10}m^{-2}$ to $10^{-15}m^{-2}$. We also obtain the compactness factor and surface redshift of the star. We observe that the compactness of the star increases as the cosmological constant decreases, whereas the surface redshift of the star decreases with a decrease in the cosmological constant. The compactness factor and surface redshift corresponding to the maximum mass of the neutron star remains almost constant as $\Lambda$ decreases.
\end{abstract}

\section{Introduction}

Einstein's General Relativity precisely describes the dynamics of planets, stars and the solar system phenomena. But the observation by Hubble concerning the accelerating universe is where Einstein's gravity needed first modification. Historically, the cosmological constant($\Lambda$) was introduced by Einstein himself to account for a static, closed, finite universe. But as observations proved universe to be expanding, $\Lambda$ was removed from the equations. Developments in particle physics accounted that the cosmological constant could be a measure of the energy density of vacuum\cite{sean}. To explain the accelerating universe, there are various modified theories of gravity such as f(R) gravity\cite{fR}, scalar-tensor theories\cite{scalartensor}, brane world cosmology\cite{braneworld} etc. The addition of the cosmological constant to Einstein's gravity is the most simplistic theory of these, which can explain the accelerated expansion. The cosmological constant is also argued to be the simplest possible form of dark energy which may be responsible for the accelerated expansion of the universe. This forms the basis of the $\Lambda$-CDM(cold dark matter) model which is in accordance with various observational results\cite{sean,neutronstarcosmo,turner}. 

Compact objects such as white dwarfs, neutron stars and black holes have considerably higher mass relative to their radii. They are formed when supermassive stars reach their last stage of evolution. Their core density is generally of the order of $10^{17}\textnormal{kg/m}^{3}$ and their radius is around $ 10^4m$ \cite{kodama,wheeler1,suwa}. Compact objects are studied using the Tolman–Oppenheimer–Volkoff (TOV) equations, which connect the dynamics of its gravitational field with the object's energy content and the latter depends on the equation of state of the matter. The TOV equations give the maximum allowed mass of the neutron star to be 2.1$M_{\odot}$ in \cite{ozel,chamel} whereas the existence of a neutron star with mass 2.16$M_{\odot}$ has been reported in \cite{most}. PSR $J0740+6620$ is a massive neutron star reported with its mass being 2.14 \(M_\odot\)\cite{ran}. Using a particular model\cite{vaidya}, the upper bound on the mass of a neutron star is found to be 3.575 $M_{\odot}$ and its generalization to a non-commutative space-time in \cite{neutronstar} found the maximum mass to be $3.6805M_{\odot}$.

Various studies of compact objects in the presence of cosmological constant have been reported in recent times. In \cite{whitedwarf}, properties of white dwarf are studied using TOV equations in Einstein-$\Lambda$ gravity. The authors showed that for $\Lambda<10^{-16}m^{-2}$, properties such as mass, radius, gravitational redshift have no significant change from Einstein gravity. But for larger $\Lambda$, the mass and radius of the white dwarf reduce. Effect of Einstein- $\Lambda$ gravity was seen to be more on low mass white dwarfs and an upper limit on $\Lambda$ is obtained to be $3 \times 10^{-14}m^{-2}$. In \cite{neutronstarcosmo}, the structure of neutron star is studied in d-dimensional Einstein-$\Lambda$ gravity using modified TOV equations and hydrostatic equilibrium equations. The maximum mass of the neutron star was found to decrease with an increase in $\Lambda$ in this model and the neutron star mass was found to satisfy the upper bound $M \leq 1.68M_{\odot}$. Significant changes in structural properties of the neutron star were only observed for $\Lambda>10^{-14} m^{-2}$. It was also reported that a negative value for the cosmological constant leads to inconsistent behavior of the variation of mass with respect to radius. In \cite{hybridneutronstar}, the structure of hybrid stars is reported to have a quark core in Einstein-$\Lambda$ gravity. For $\Lambda > 10^{-14}m^{-2}$, the maximum mass and radius of the hybrid star decrease with an increase in $\Lambda$. As the gravitational redshift of compact objects cannot be greater than unity, an upper bound on $\Lambda$ is obtained at $9 \times 10^{-13}m^{-2}$. In \cite{garattini}, modifications of TOV equations are derived and analyzed for rainbow gravity. Magnetic neutron stars in an energy-dependent metric are studied in \cite{panah}. The maximum mass and radius of the star are found to increase with the magnetic field, whereas average density and redshift factor decrease with an increase in magnetic field. These authors also showed that magnetized neutron stars can have a mass greater than $3.2M_{\odot}$ in rainbow gravity. Modifications to neutron star properties are studied in Rastall-Rainbow gravity \cite{mota}, where the mass of the star is found to decrease with an increase in its radius by fixing either the rainbow parameter or the Rastall parameter. In \cite{ruff}, the maximum allowed mass for a neutron star is found to be $3.2M_{\odot}$ using the functional maximization procedure with physical constraints. In \cite{hendi}, the hydrostatic equilibrium equation of neutron stars is derived and studied from a spherically symmetric (3+1)-dimensional metric with the cosmological constant in rainbow's gravity. The mass of the star was found to increase/decrease with the rainbow parameter depending on the value of the parameter being greater than/less than unity. In this case, the maximum mass of the neutron star is found to be $2.81M_{\odot}$. Modified hydrostatic equilibrium equations in gravity's rainbow are derived and studied for anisotropic fluid given by extended Chaplygin equation in \cite{tudeshki}. The maximum mass of the dark energy star in rainbow gravity was found to be $3.70M_{\odot}$ in the isotropic model and $3.90M_{\odot}$ in the anisotropic model. As the compact objects posses large gravitational field, it serves as a testing ground for models studying quantum gravity effects.   


 Doubly special relativity(DSR) incorporates two fundamental constants, namely the speed of light and the Planck length in such a way that they are compatible with the deformed relativity principle. Incorporating general relativity with doubly special relativity gives rainbow gravity where the geometry of the space-time is considered to be dependent on the energy of the particle traversing through it\cite{rainbowgravity}. It provides a UV completion theory as Einstein's gravity has fundamental problems in the UV limit and reduces to the classical theory in the IR limit. In \cite{blackholeali}, black hole physics is investigated in rainbow gravity and corrections to temperature and entropy of the black hole due to rainbow gravity modifications are shown to prevent the evaporation of black hole completely, leaving remnant black holes. 
 
The space-time associated with doubly special relativity is the $\kappa$- deformed space-time. The time and space coordinates of this space-time \cite{kappa1, dimitrijevic, dasz, mel1} do not commute whereas the space coordinates commute among themselves, i.e., 
\begin{equation}\label{ksp-1}
[\hat{x}^i,\hat{x}^j]=0,~~~[\hat{x}^0, \hat{x}^i]=ia\hat{x}^i,~~~a=\frac{1}{\kappa}.
\end{equation}
Here, $a$ is the deformation parameter with the length dimension. In the limit $a\rightarrow 0$ one gets the commutative space-time.
An unavoidable consequence of the non-commutativity of space-time is the introduction of a fundamental length scale. As fundamental length scale is intrinsic to all quantum gravity models, the appearance of a natural length scale in non-commutative space-time makes it suitable to model quantum gravity effects.

In this paper, we construct and study neutron star in the presence of cosmological constant in the $\kappa$-deformed space-time. For this, we adapt the approach of \cite{vaidya} to model a neutron star in the $\kappa$-deformed space-time with the cosmological constant. We derive the corresponding TOV equations and obtain their solutions. We further analyze these solutions and discuss the salient features of our neutron star solution.

In recent times, different aspects of black hole physics in $\kappa$-deformed space-time have been studied \cite{zuhair1,kappa-btz,kappa-btz1,gupta5,mel3}. Compact stars in the $\kappa$-deformed space-time were investigated in \cite{zuhair2,vishnu}. Evolution of the universe using Newtonian cosmology is studied in $\kappa$-deformed space-time \cite{suman}. In \cite{bani}, the super Chandrasekhar limit is obtained using a squashed fuzzy sphere. In \cite{neutronstar}, neutron star in $\kappa$-deformed space-time was studied and the maximum mass of the neutron star is found to be $3.6805M_{\odot}$. 

In this paper, we investigate the modifications to neutron star parameters in $\kappa$-deformed space-time. We start by generalizing the metric
\begin{equation}
ds^2=e^{\nu(r)}dt^2-\frac{1-K\frac{r^2}{R^2}}{1-\frac{r^2}{R^2}}dr^2-r^2\left(d\alpha^2+sin^2\alpha d\beta^2\right).
\label{metric}
\end{equation}
to $\kappa$-deformed space-time. In the above equation, $K=1-\frac{l^2}{R^2}$, $R$ denotes the equatorial radius of the $3$-spheroid, and $l$ denotes the distance from the center to the pole along the positive symmetry axis. We analyze the neutron star model in the presence of the cosmological constant in $\kappa$-deformed space-time. The non-commutative metric and energy-momentum tensor are constructed and using these we set up TOV equations valid up to first order in $a$. By solving these TOV equations we find the $\kappa$-deformed mass and radius of the neutron star in the presence of the cosmological constant. We observe that for a fixed value of the deformation parameter $a$, the mass and radius of the star decrease as $\Lambda$ decreases.


This paper is organized as follows. In Section 2, we give a brief introduction to $\kappa$-deformed space-time. In Section 3, we construct the $\kappa$-deformed field equations in the presence of the cosmological constant and derive the TOV equations. These field equations are solved in Section 4, which gives mass, radius, compactness factor and surface redshift of the neutron star in $\kappa$-deformed space-time, valid up to first order in $a$ with the corrections due to the cosmological constant. Section 5 presents the results and concluding remarks.

\section{Kappa Deformed Space-Time}
Field theoretical models in $\kappa$-deformed space-time are constructed by either using the star product formalism\cite{dimitrijevic,dasz} or using the realization approach\cite{mel1,mel2}. Since these two methods are equivalent, the calculations in this paper are done by adopting the realization method. In this approach, we re-express the non-commutative coordinates using commutative coordinates and their derivatives.
The $\kappa$-deformed coordinate, $\hat{x}_{\mu}$ is written as \cite{mel1}
\begin{equation}\label{ksp-2}
\begin{split}
 \hat{x}_0=&x_0\psi(ia\partial_{0})+iax_j\partial_j\gamma(ia\partial_{0})\\
 \hat{x}_i=&x_i\varphi(ia\partial_{0}).
\end{split}
\end{equation}
Here $\psi$, $\gamma$, and $\varphi$ are functions of $ia\partial_{0}$ and satisfy the conditions
\begin{equation}\label{ksp-3}
 \psi(0)=1,~\varphi(0)=1.
\end{equation}
Substituting eq.(\ref{ksp-2}) in eq.(\ref{ksp-1}), we see that
\begin{equation}\label{ksp-4}
 \frac{\varphi'(ia\partial_{0})}{\varphi(ia\partial_{0})}\psi(ia\partial_{0})=\gamma(ia\partial_{0})-1,
\end{equation}
where prime denotes differentiation with respect to $ia\partial_{0}$. $\psi(ia\partial_{0})$ can take the values $1$ or $1+2(ia\partial_{0})$  \cite{mel1}. In our study, we choose $\psi(ia\partial_{0})=1$ which leads to equations (\ref{ksp-2}) taking the form
\begin{equation}\label{ksp-5}
\begin{split}
 \hat{x}_0=&x_0+iax_j\partial_j\gamma(ia\partial_{0})\\
 \hat{x}_i=&x_i\varphi(ia\partial_{0}),
\end{split}
\end{equation}
 The allowed choices for $\varphi$ are $e^{-ia\partial_{0}}, e^{-\frac{ia\partial_{0}}{2}}, 1, \frac{ia\partial_{0}}{e^{ia\partial_{0}}-1}$, etc. The generalized form of free particle dispersion relation is given by \cite{mel1}
\begin{equation}
\frac{4}{a^2}\sinh^2 \bigg(\frac{ap^{0}}{2}\bigg) -p_ip_i \frac{e^{-ap^{0}}}{\varphi^2(ap^{0})}-m^2c^2 +\frac{(ap^{0})^2}{4}\left[\frac{4}{a^2}\sinh^2\bigg(\frac{ap^{0}}{2}\bigg) -p_ip_i \frac{e^{-ap^{0}}}{\varphi^2(ap^{0})}\right]^2=0.\label{disp}
\end{equation}
Here, $p^i$ is the component of the commutative 3-momenta of the particle. In the limit $a\rightarrow 0$, the commutative energy-momentum relation is obtained. 
 Using the realization $\varphi(ia\partial_{0})=e^{-ap^{0}}$, from the above equation (eq.(\ref{disp})), we find \cite{neutronstar}
\begin{equation}
\frac{\hat{E}}{c}=\frac{E}{c}\left(1+\frac{1}{2} a \frac{p_i^2 c}{E}\right),
\label{E_correction}
\end{equation}
where $\hat E$ represents the non-commutative energy and $p^{0}=\frac{E}{c}$. We re-write the above equation as
\begin{equation}
\hat{E}=E\left[1+\frac{1}{2}a p^0\left\{1-\left(\frac{m c^2}{E}\right)^2\right\}\right]\equiv Eg(E).
\label{dispersion_relation}
\end{equation}
\section{$\kappa$-deformed Einstein's field equation with the cosmological constant}

Here, we first construct non-commutative metric in $\kappa$-deformed space-time describing the interior of the neutron star. For this, we start with the generalized commutation relation of $\kappa$-deformed phase space coordinates given by \cite{kappa-geod},
\begin{equation}\label{N1}
 [\hat{x}_{\mu},\hat{P}_{\nu}]=i\hat{g}_{\mu\nu}, 
\end{equation} 
where $\hat{g}_{\mu\nu}(\hat{x}^{\alpha})$ denotes the $\kappa$-deformed metric. The $\kappa$-deformed phase-space coordinates can be written as \cite{kappa-geod},
\begin{equation}\label{N2}
 \hat{x}_{\mu}=x_{\alpha}\varphi^{\alpha}_{\mu},~~ \,\hat{P}_{\mu}=g_{\alpha\beta}(\hat{y})p^{\beta}\varphi^{\alpha}_{\mu},
\end{equation}
where $\hat{P}_{\mu}$ is the $\kappa$-deformed generalized momenta corresponding to the non-commutative coordinate $\hat{x}_{\mu}$ and $p_{\mu}$ is the canonical conjugate momenta corresponding to the commutative coordinate $x_{\mu}$. We see that in the limit, i.e., $a\to 0$ (from eq.(\ref{N2})), we get the corresponding commutative coordinate $x_{\mu}$ and commutative momenta $p_{\mu}$. The coordinates $\hat{y}_{\mu}$ introduced in the above equation are also assumed to satisfy the commutation relations,
 
\begin{equation}
[\hat{y}_i,\hat{y}_j]=0~~~~[\hat{y}_0,\hat{y}_i]=ia\hat{y}_i \label{N3b}
\end{equation}
Also, $[\hat{y}_{\mu},\hat{x}_{\nu}]=0$. They are introduced for calculational simplification \cite{kappa-geod}. The functional form of $g_{\alpha\beta}(\hat{y})$ in eq.(\ref{N2}) is the same as that of the metric in the commutative space-time except that $x_{\mu}$ is replaced with non-commutative coordinate $\hat{y}_{\mu}$.

Substituting eq.(\ref{N2}) in eq.(\ref{N1}), we obtain $\varphi_{\mu}^{\alpha}$ as
\begin{equation}\label{N3}
 \varphi _0^0=1, \, \varphi _i^0=0, \, \varphi_0^i=0, \, \varphi _j^i=\delta _j^i e^{-ap^0}. 
\end{equation}
We express $\hat{y}_{\mu}$ in terms of commutative coordinates and corresponding momenta as, 
\begin{equation}\label{N4}
 \hat{y}_{\mu}=x_{\alpha}\phi_{\mu}^{\alpha}.
 \end{equation}
Using eq.(\ref{N3b}) and $[\hat{x}_{\mu},\hat{y}_{\nu}]=0$, we get $\phi_{\mu}^{\alpha}$ as (for details see \cite{kappa-geod})
\begin{equation}\label{N5}
 \phi_{0}^{0}=1,~\phi_{i}^{0}=0,~\phi_{0}^{i}=-ap^i,~\phi_{i}^{j}=\delta_{i}^{j}.
\end{equation}
which gives the explicit form of $\hat{y}_{\mu}$ as
\begin{equation}\label{N6}
 \hat{y}_0=x_0-ax_jp^j,~~
\hat{y}_i=x_i.
 \end{equation}
Using the above in eq.(\ref{N2}) and substituting $\hat{x}_{\mu}$ and $\hat{P}_{\mu}$ in eq.(\ref{N1}), we get \cite{kappa-geod}
\begin{equation}\label{N7}
 [\hat{x}_{\mu},\hat{P}_{\nu}] \equiv i\hat{g}_{\mu\nu}=ig_{\alpha\beta}(\hat{y})\Big(p^{\beta}\frac{\partial \varphi^{\alpha}_{\nu}}{\partial p^{\sigma}}\varphi_{\mu}^{\sigma}+\varphi_{\mu}^{\alpha}\varphi_{\nu}^{\beta}\Big). \end{equation}


Substituting eq.(\ref{N3}) in eq.(\ref{N7}), the explicit form of the components of $\hat{g}_{\mu\nu}$ is found to be
\begin{equation}\label{N9}
\begin{aligned}
\hat{g}_{00}&=g_{00}(\hat{y}),\\
\hat{g}_{0i}&=g_{0i}(\hat{y})\big(1-ap^0\big)e^{-ap^0}-ag_{im}(\hat{y})p^me^{-ap^0},\\ 
\hat{g}_{i0}&=g_{i0}(\hat{y})e^{-ap^0},\\
\hat{g}_{ij}&=g_{ij}(\hat{y})e^{-2ap^0}.
\end{aligned}
\end{equation}
Since eq.(\ref{N6}) indicates that the metric components do not have explicit time dependence, we find $g_{\mu\nu}(\hat{y})=g_{\mu\nu}(x^i)$. As the metric in eq.(\ref{metric}) is spherically symmetric, the $\kappa$ deformed metric takes the form\footnote{$\kappa$-deformed space-time is invariant under rotations. In deriving $\kappa$-deformed metric, we use realization(see eq.(\ref{N6})) for non-commutative coordinate $\hat{y}_{\mu}$ for calculational simplification. Due to this realization, we get off-diagonal terms in $\kappa$-deformed metric. Thus to retain rotational symmetry, the off-diagonal terms in the $\kappa$-deformed metric are set to be zero.},
\begin{equation}\label{N12}
 d\hat{s}^2=g_{00}(\hat{y})dx^0dx^0+g_{ij}(\hat{y})e^{-4ap^0}dx^idx^j. 
\end{equation}
 Thus the $\kappa$-deformed generalisation of the metric given in eq.(\ref{metric}) is
\begin{equation}
d\hat{s}^2=e^{\nu(r)}dt^2-e^{-4ap^0}\left[\frac{1-K\frac{r^2}{R^2}}{1-\frac{r^2}{R^2}}dr^2+r^2\left(d\alpha^2+sin^2\alpha d\beta^2\right)\right].
\label{kdeformed_metric}
\end{equation}
The generalised energy-momentum tensor in $\kappa$-deformed space-time is \cite{neutronstar}
\begin{equation}
\hat{T}^{\alpha\beta}=e^{3ap^0}g(E)\varphi^{\alpha}_{\sigma}\varphi^{\beta}_{\delta}T^{\sigma\delta}.
\label{energy_momentum_tensor2}
\end{equation}
We know that in the proper frame, $T$ is $diag(\rho c^2, P, P, P)$, where $\rho$ and P are the density and pressure of the fluid, respectively. Using these in eq.(\ref{energy_momentum_tensor2}) we find,
\begin{eqnarray}
\hat{T}^{00}=e^{3ap^0}g(E)\rho c^2\equiv \hat{\rho}c^2\nonumber\\
\hat{T}^{ij}=e^{ap^0}g(E)\delta_{ij}P\equiv \hat{P}\delta_{ij}.
\label{hat_rho_p}
\end{eqnarray}
We get the deformed density and pressure in $\kappa$-deformed space-time as $\hat{\rho}=e^{3ap^0}g(E)\rho $ and $\hat{P}=e^{ap^0}g(E)P$, respectively. Thus the general form of the deformed energy-momentum tensor for a fluid is
\begin{equation}
\hat{T}^{\mu\nu}=\left(\hat{\rho}+\frac{\hat{P}}{c^2}\right)\hat{u}^{\mu}\hat{u}^{\nu}-\hat{P}\hat{g}^{\mu\nu},
\label{energy_momentum_tensor}
\end{equation}
where $\hat{u}^{\mu}=\frac{d\hat{x}^{\mu}}{d\hat{\tau}}$. To find the explicit form of  $\hat{T}_{\mu\nu}$ valid for the neutron star, we assume the interior of the star to be a static perfect fluid and $\hat{u}^{\mu}=(c e^{-\nu(r)/2},0,0,0)$. Thus the non-vanishing components of eq.(\ref{energy_momentum_tensor}) become
\begin{eqnarray}
\hat{T}_{00}&=&c^2 e^{\nu(r)}\hat{\rho}\nonumber\\
\hat{T}_{11}&=&\hat{P}\frac{1-K\frac{r^2}{R^2}}{1-\frac{r^2}{R^2}}e^{-4 a p^{0}}\nonumber\\
\hat{T}_{22}&=&\hat{P}r^2 e^{-4 a p^{0}}\nonumber\\
\hat{T}_{33}&=&\hat{P}r^2 sin^2\alpha e^{-4 a p^{0}}.
\label{component_em_tensor}
\end{eqnarray}
Now, the $\kappa$-deformed Einstein's field equation with the cosmological constant takes the form,
\begin{equation}
\frac{8\pi G}{c^4}\hat{T}_{\mu\nu}=\hat{R}_{\mu\nu}-\frac{1}{2}\hat{R}\hat{g}_{\mu\nu}+\hat{\Lambda}\hat{g_{\mu\nu}},
\label{Einstein_field_equation(cosmological constant)}
\end{equation}
where $\hat{\Lambda}=\Lambda(1+ap^{0})$ is the $\kappa$-deformed cosmological constant and $\hat{R}_{\mu\nu}$, $\hat{R}$ represent the $\kappa$-deformed Ricci tensor and Ricci scalar, respectively. Using expressions of $\hat{R}_{\mu\nu}$ and $\hat{R}$ (see for details \cite{neutronstar}) and $\hat{T}_{\mu\nu}$ given in eq.(\ref{component_em_tensor}) in eq.(\ref{Einstein_field_equation(cosmological constant)}), we find the field equations to be
\begin{equation}
\frac{8\pi G}{c^2}\hat{\rho}=e^{4 ap^0}\frac{3(1-K)}{R^2}\frac{\left(1-K\frac{r^2}{3R^2}\right)}{\left(1-K\frac{r^2}{R^2}\right)^2}+\hat{\Lambda}
\label{00}
\end{equation}
\begin{equation}
\frac{8\pi G}{c^4}\hat{P}=e^{4ap^0}\left\{\Bigg(\frac{1-\frac{r^2}{R^2}}{1-K\frac{r^2}{R^2}}\Bigg)\left(\frac{\nu^{\prime}(r)}{r}+\frac{1}{r^2}\right)-\frac{1}{r^2}\right\}-\hat{\Lambda}
\label{11}
\end{equation}
\begin{eqnarray}
\left(1-\frac{r^2}{R^2}\right)\left(1-K\frac{r^2}{R^2}\right)\left\{\nu^{\prime\prime}(r)+\frac{1}{2}\left[\nu^{\prime}(r)\right]^2-\frac{\nu^{\prime}(r)}{r}\right\}\nonumber\\
-\frac{2(1-K)}{R^2}r\left[\frac{\nu^{\prime}(r)}{2}+\frac{1}{r}\right]+\frac{2(1-K)}{R^2}\left(1-K\frac{r^2}{R^2}\right)=0,
\label{22}
\end{eqnarray}
 where $\nu^{\prime}(r)=\frac{d\nu}{dr}$ and $\nu^{\prime\prime}(r)=\frac{d^2\nu}{dr^2}$. The $\kappa$- deformed field equations with the cosmological constant obtained in eq.(\ref{00}-\ref{22}) reduce to the corresponding results obtained in \cite{neutronstar} by setting $\hat{\Lambda}=0$. The corresponding commutative result is obtained in the limit $a\rightarrow 0$. For $r=0$ in eq.(\ref{00}), we have
\begin{equation}
\frac{8\pi G}{c^2}\hat{\rho}_{0}=e^{4ap^0}\frac{3(1-K)}{R^2}+\hat{\Lambda}
\label{rho_r=0}
\end{equation}
On the boundary of the star where $r=b$ ($b$ being the radius of the star), we have
\begin{equation}
\frac{8\pi G}{c^2}\hat{\rho}_{b}=e^{4ap^0}\frac{3(1-K)}{R^2}\frac{\left(1-K\frac{b^2}{3R^2}\right)}{\left(1-K\frac{b^2}{R^2}\right)^2}+\hat{\Lambda}.
\label{rho_r=b}
\end{equation}
Taking derivative of eq.(\ref{00}) with respect to $r$ and substituting for $\hat{\rho}$ from eq.(\ref{hat_rho_p}) we get,
\begin{equation}
\frac{8\pi G}{c^2}g(E)\frac{d\rho}{dr}=e^{ap^0}\frac{10K(1-K)r}{R^4}\frac{\left(1-\frac{K r^2}{5R^2}\right)}{\left(1-K\frac{r^2}{R^2}\right)^3}~.
\label{derivative_rho}
\end{equation}
From eq.(\ref{derivative_rho}), we see that, since $\rho$ decreases with an increase in $r$ and is always positive, $K$ has to be negative, i.e., $K<0$. 

We know that the ratio of the density of the star at the boundary to that at the center ($\lambda=\frac{\rho_b}{\rho_0}$) is always less than $1$, which gives from eq.(\ref{rho_r=0}) and eq.(\ref{rho_r=b})
\begin{eqnarray}
\lambda=\frac{\rho_b}{\rho_0}=\frac{\frac{1-K\frac{b^2}{3 R^2}}{\left(1-K\frac{b^2}{R^2}\right)^2}+\Sigma}{1+\Sigma}<1,
\label{def_lambda}
\end{eqnarray}
where $\Sigma=\hat{\Lambda}e^{-4ap^{0}}\frac{R^2}{3(1-K)}$. Solving the above, we find

\begin{eqnarray}
\frac{b^2}{R^2}=\frac{6\lambda-1+6\Sigma(\lambda-1)-\sqrt{1+24\lambda-24\Sigma(1-\lambda)}}{6 K\left[\lambda+\Sigma(\lambda-1)\right]}.
\label{b/R}
\end{eqnarray}
The space-time exterior to the neutron star is described by the $\kappa$- deformed generalization of the de-Sitter metric \cite{desitter} given by 
\begin{equation}
ds^2=\left(1-\frac{2 M}{r}-\frac{\hat{\Lambda}r^2}{3}\right)dt^2-e^{-4ap^0}\left[\frac{dr^2}{1-\frac{2 M}{r}-\frac{\hat{\Lambda}r^2}{3}}+r^2\left(d\alpha^2+sin^2\alpha d\beta^2\right)\right].
\label{deSitter}
\end{equation}
As $g_{11}$ has to be continous at the boundary $r=b$, from eq.(\ref{kdeformed_metric}) and eq.(\ref{deSitter}) we have
\begin{eqnarray}
 M=\frac{b^3(1-K)}{2R^2\left(1-K\frac{b^2}{R^2}\right)}-\frac{\hat{\Lambda}b^3}{6}.
\label{mass}
\end{eqnarray}
From eq.(\ref{rho_r=0}), we find
\begin{equation}
R=\sqrt{\frac{e^{ap^0}3(1-K)c^2}{8\pi G \rho_{0}g(E)-\hat{\Lambda}e^{-3ap^0}}}.
\label{radius}
\end{equation}
 The mass and radius of the star can be obtained by setting values for $\rho_{b},\lambda,K,ap^0,\frac{mc^2}{E}$ and substituting them in eq.(\ref{b/R}), eq.(\ref{mass}) and eq.(\ref{radius}). For this we need to generalize strong energy and causality conditions in $\kappa$-deformed space-time for which solutions to eq.(\ref{11}) and eq.(\ref{22}) are required.

\section{Properties of the $\kappa$-deformed neutron star with the cosmological constant}
We now analyze the modification of various parameters associated with the $\kappa$-deformed neutron star when the cosmological constant is non-zero. For this, we need to solve TOV equations(see eq.(\ref{00},\ref{11},\ref{22})). We first obtain $\nu(r)$ by solving eq.(\ref{22}) using series solution method and the solutions of eq.(\ref{00}) and eq.(\ref{11}) along with the causality condition(see eq.(\ref{causality}) below) allow us to calculate permitted values of the mass, radius, compactness factor and surface redshift of the neutron star. This $\nu(r)$ is then substituted back in eq(\ref{00}) and eq.(\ref{11}). Since eq.(\ref{22}) is a second-order differential equation, its solution involves two arbitrary constants. We use the boundary conditions and fix these constants in terms of $M$ and $R$ given in eq.(\ref{mass}) and eq.(\ref{radius}), respectively.  


To solve the eq.(\ref{22}) for $\nu(r)$, we make the change of variables
\begin{eqnarray}
\psi=e^{\nu/2}; ~~ u=\sqrt{\frac{K}{K-1}}\sqrt{1-\frac{r^2}{R^2}}.
\label{variable_change}
\end{eqnarray}
which reduces eq.(\ref{22}) to 
\begin{equation}
(1-u^2)\frac{d^2\psi}{du^2}+u\frac{d\psi}{du}+(1-K)\psi=0.
\label{change22}
\end{equation}
Substituting $\psi=\sum_{n=0}^{\infty}a_{n}u^n$ in eq.(\ref{change22}), we get the recursion relation,
\begin{equation}
(n+1)(n+2)a_{n+2}=(n^2-2n+K-1)a_n.
\label{recursion}
\end{equation}
For terminating the series, we choose $K$ that satisfies
\begin{equation}
n^2-2n+K-1=0,~~which~gives~~ n=1\pm\sqrt{2-K}.
\label{condition}
\end{equation}
In order to terminate the series, the simplest choice for $K$ is $K=-2$. Thus, we find
\begin{equation}
e^{\frac{\nu}{2}}=Bd_{e}\left(1-\frac{4}{9}d_{e}^2\right)+C\left(1-\frac{2}{3}d_{e}^2\right)^\frac{3}{2},
\end{equation}
where $d_{e}=\sqrt{1-\frac{r^2}{R^2}}$. As $r$ varies from 0 to $R$(equatorial radius of the star), $d_{e}$ decreases from 1 to 0. Solution to $\nu(r)$ gives the final form of the deformed metric as
\begin{eqnarray}
d\hat{s}^2&=&\left\{Bd_{e}\left(1-\frac{4}{9}d_{e}^2\right)+C\left(1-\frac{2}{3}d_{e}^2\right)^\frac{3}{2}\right\}^2dt^2\nonumber\\
&-&e^{-4ap^0}\left[\frac{3-2d_{e}^2}{d_{e}^2}dr^2+r^2\left(d\alpha^2+sin^2\alpha d\beta^2\right)\right].
\label{final_metric}
\end{eqnarray}
 Solution for $\nu(r)$ is substituted in eq.(\ref{00}) and eq.(\ref{11}) with $K=-2$ gives,
\begin{equation}
\frac{8\pi G}{3 c^2}\hat{\rho}=e^{4 ap^0}\frac{5-2 d_{e}^2}{R^2\left(3-2 d_{e}^2\right)^2}+\frac{\hat{\Lambda}}{3}
\label{rho_z}
\end{equation}
\begin{equation}
\frac{8 \pi G}{c^4}\hat{P}=e^{4 ap^0}\frac{3}{R^2}\left\{\frac{C(2d_{e}^2-1)\left(1-\frac{2}{3}d_{e}^2\right)^\frac{1}{2}-\frac{1}{3}B d_{e}(5-4d_{e}^2)}{(3-2d_{e}^2)\left[C\left(1-\frac{2}{3}d_{e}^2\right)^\frac{3}{2}+B d_{e}\left(1-\frac{4}{9}d_{e}^2\right)\right]}\right\}-\hat{\Lambda}
\label{P_z}
\end{equation}
The explicit form of the metric (eq.(\ref{final_metric})) and the subsequent equations(eq.(\ref{rho_z}),eq.(\ref{P_z})) are valid in the interior of the star. Hence, we match the exterior de-Sitter metric(eq.(\ref{deSitter})) to the final form of the neutron star metric (eq.(\ref{final_metric})) at the boundary $r=b$ which gives 
\begin{equation}
\left(1-\frac{2M}{b}-\frac{\hat{\Lambda}b^2}{3}\right)=\frac{1-\frac{b^2}{R^2}}{1+2\frac{b^2}{R^2}}
\label{g11}
\end{equation}
\begin{equation}
B\left(1-\frac{b^2}{R^2}\right)^{\frac{1}{2}}\left(5+4\frac{b^2}{R^2}\right)+C\sqrt{3}\left(1+2\frac{b^2}{R^2}\right)^{\frac{3}{2}}=9\left(1-\frac{2M}{b}-\frac{\hat{\Lambda} b^2}{3}\right)^{\frac{1}{2}}.
\label{g00}
\end{equation}
At the boundary, the fluid pressure must vanish \cite{vaidya} and substituting for $d_{e}$ in eq.(\ref{P_z}), we find
\begin{eqnarray}
B\left(1-\frac{b^2}{R^2}\right)^\frac{1}{2}\left(1+4\frac{b^2}{R^2}\right)+\frac{\hat{\Lambda}R^2}{9}e^{-4ap^0}B\left(1+2\frac{b^2}{R^2}\right)\left(1-\frac{b^2}{R^2}\right)^\frac{1}{2}\left(5+4\frac{b^2}{R^2}\right)   \nonumber \\ 
=C\sqrt{3}\left(1-2\frac{b^2}{R^2}\right)\left(1+2\frac{b^2}{R^2}\right)^\frac{1}{2}-\frac{\hat{\Lambda}R^2}{9}e^{-4ap^0}\sqrt{3}C\left(1+2\frac{b^2}{R^2}\right)\left(1+2\frac{b^2}{R^2}\right)^\frac{3}{2}.
\label{pressurecondition}
\end{eqnarray}
Solving eq.(\ref{g11}), eq.(\ref{g00}) and eq.(\ref{pressurecondition}) we get,
\begin{eqnarray}
B&=&\frac{3}{2}\frac{1-2\frac{b^2}{R^2}}{\left(1+2\frac{b^2}{R^2}\right)^\frac{1}{2}}-\frac{\hat{\Lambda}}{6}R^2 e^{-4ap^0}\left(1+2\frac{b^2}{R^2}\right)^\frac{3}{2}\nonumber\\
C&=&\frac{\sqrt{3}}{2}\sqrt{1-\frac{b^2}{R^2}}\left\{\frac{1+4\frac{b^2}{R^2}}{1+2\frac{b^2}{R^2}}\right\}+\frac{\hat{\Lambda}e^{-4ap^0}R^2}{6\sqrt{3}}\left\{\frac{\sqrt{1-\frac{b^2}{R^2}}}{1+2\frac{b^2}{R^2}}\right\}\left[8\frac{b^4}{R^4}+14\frac{b^2}{R^2}+5\right]
\label{AB}
\end{eqnarray}
Dividing eq.(\ref{P_z}) by eq.(\ref{rho_z}), we find
\begin{multline}
\frac{\hat{P}}{\frac{1}{3}\hat{\rho}c^2}=\frac{e^{4ap^0}3(3-2d_{e}^2)\left[C(2d_{e}^2-1)\left(1-\frac{2}{3}d_{e}^2\right)^\frac{1}{2}-\frac{Bd_{e}}{3}(5-4d_{e}^2)\right]}{\left[e^{4ap^0}(5-2d_{e}^2)+\frac{\hat{\Lambda}R^2}{3}(3-2d_{e}^2)^2\right]\left[C\left(1-\frac{2}{3}d_{e}^2\right)^\frac{3}{2}+B d_{e}\left(1-\frac{4}{9}d_{e}^2\right)\right]}-  \\ 
\frac{\hat{\Lambda}(3-2d_{e}^2)^2\left[Bd_{e}\left(1-\frac{4}{9}d_{e}^2\right)+C\left(1-\frac{2}{3}d_{e}^2\right)^\frac{3}{2}\right] }{\left[e^{4ap^0}(5-2d_{e}^2)+\frac{\hat{\Lambda}R^2}{3}(3-2d_{e}^2)^2\right]\left[C\left(1-\frac{2}{3}d_{e}^2\right)^\frac{3}{2}+B d_{e}\left(1-\frac{4}{9}d_{e}^2\right)\right]}
\label{P/rho}
\end{multline}
Now we impose the $\kappa$- deformed strong energy condition given by,
\begin{equation}
\hat{P}<\frac{1}{3}\hat{\rho}c^2.
\label{strong_energy}
\end{equation}
The above strong energy condition is satisfied if $0<\frac{\hat{P}}{\frac{1}{3}\hat{\rho}c^2}<1$ and it is verified by substituting eq.(\ref{AB}) and eq.(\ref{P/rho}) in the inequality. We see that as $d_{e}$ changes from 1 to 0, we move from the center of the star to its boundary. Thus we get an upper bound for $\frac{b^2}{R^2}$ if the strong energy condition is satisfied for $d_{e}=1$, i.e., for the strong energy condition to be satisfied everywhere in the star, we get a bound on $\frac{b^2}{R^2}$ at $d_{e}=1$. Here, the upper bound varies with choice of $\lambda$ as eq.(\ref{AB}) has $R$ dependence which in-turn has implicit $\lambda$ dependence through $\rho_{0}$(eq.(\ref{radius})). 

Condition of causality that our solution needs to satisfy sets the speed of sound to be always less than the speed of light. We have speed of sound given by $\sqrt{\frac{dP}{d\rho}}$ and hence the causality condition in $\kappa$-deformed space-time takes the form,

\begin{equation}
\frac{d\hat{P}}{d\hat{\rho}}<c^2.
\label{causality}
\end{equation}
In order to verify that eq.(\ref{causality}) is satisfied, we take the derivatives of eq.(\ref{rho_z}) and eq.(\ref{P_z}) with respect to $d_{e}$, which gives

\begin{equation}
\frac{1}{c^2}\frac{d\hat{P}}{d\hat{\rho}}=\frac{1}{\frac{d}{d(d_{e})}\left\{\frac{5-2 d_{e}^2}{\left(3-2 d_{e}^2\right)^2}\right\}}\frac{d}{d(d_{e})}\left\{\frac{(2d_{e}^2-1)\left(1-\frac{2}{3}d_{e}^2\right)^\frac{1}{2}-\frac{1}{3}\frac{B}{C} d_{e}(5-4d_{e}^2)}{(3-2d_{e}^2)\left[\left(1-\frac{2}{3}d_{e}^2\right)^\frac{3}{2}+\frac{B}{C} d_{e}\left(1-\frac{4}{9}d_{e}^2\right)\right]}\right\}
\end{equation}

\begin{table}[h!]
\caption{Condition on the value of $\frac{B}{C}$ for different values of $d_{e}=\sqrt{1-\frac{r^2}{R^2}}$ to satisfy the causality condition.\label{tab1}}
\centering   
\begin{tabular}{|c|c|}
\hline
\textbf{$d_{e}=\sqrt{1-\frac{r^2}{R^2}}$}& \textbf{Condition on the value of $\frac{B}{C}\equiv  x$ due to constraint $0<\frac{1}{c^2}\frac{d\hat{P}}{d\hat{\rho}}<1$} \\
    \hline
    1 & $x>0.1762$  \\
      \hline
    0.98 & $x>0.1765$ \\
      \hline
    0.96 & $x>0.1741$  \\
      \hline
    0.94 & $x>0.1691$  \\
      \hline
    0.92 & $x>0.1619$ \\
      \hline
    0.90 & $x>0.1527$ \\
     \hline
    0.88 & $x>0.1417$ \\
      \hline
   \end{tabular}
 \end{table}
 
 From Table-\ref{tab1}, we see that a bound on $\frac{B}{C}$ is obtained when the causality condition is satisfied. Since causality has to be satisfied everywhere inside the star, the lower bound for $\frac{B}{C}$ is 0.1762 corresponding to $d_{e}=1$. 
In  order to find the radius of the star, we obtain $g(E)$ from eq.(\ref{dispersion_relation}) and substitute in eq.(\ref{radius}), which gives
\begin{equation}
R=\frac{R_0}{\left(1-\frac{\hat{\Lambda}c^2}{16\pi G\rho_{0}}\right)}\left[1+ap^{0}\left\{\frac{\alpha}{2}-\frac{\beta \frac{\Lambda c^2}{16\pi G\rho_{0}}}{\left(1-\frac{\Lambda c^2}{16\pi G\rho_{0}}\right)}\right\}\right],\label{radius_final}
\end{equation}
where
\begin{equation}
R_{0}=\sqrt{\frac{9c^2}{8\pi G \rho_{0}}}~;~
\alpha\equiv 1-\frac{1}{2}\left\{1-\left(\frac{m c^2}{E}\right)^2\right\}~;~
\beta=3+\frac{1}{2}\left\{1-\left(\frac{m c^2}{E}\right)^2\right\}.
\label{radius_0}
\end{equation}
Note that we have included a correction term valid up to the first order in the deformation parameter. The radius of the star can be obtained from eq.(\ref{b/R})
\begin{equation}
b=\frac{b_0}{\left(1-\frac{\hat{\Lambda}c^2}{16\pi G\rho_{0}}\right)}\left\{1+ap^{0}\left(\frac{\alpha}{2}-\frac{\beta \frac{\Lambda c^2}{16\pi G\rho_{0}}}{\left(1-\frac{\Lambda c^2}{16\pi G\rho_{0}}\right)}\right)\right\}, \label{b_final}
\end{equation}
where
\begin{equation}
b_0=\sqrt{\frac{R_0^2}{12}\left\{\frac{1-6\lambda-6\Sigma(\lambda-1)+\sqrt{1+24\lambda-24\Sigma(1-\lambda)}}{\lambda+\Sigma(\lambda-1)}\right\}}
\end{equation}
and $\lambda$ is given in eq.(\ref{def_lambda}). To determine the mass of the star, use  eq.(\ref{radius_final}), eq.(\ref{b_final}) and eq.(\ref{mass}) and find,
\begin{eqnarray}
M&=&\frac{M_0}{\left(1-\frac{\hat{\Lambda}c^2}{16\pi G\rho_{0}}\right)}\left\{1+ap^{0}\left(\frac{\alpha}{2}-\frac{\beta \frac{\Lambda c^2}{16\pi G\rho_{0}}}{\left(1-\frac{\Lambda c^2}{16\pi G\rho_{0}}\right)}\right)\right\}\\&-&\frac{\hat{\Lambda} b_{0}^3}{6\left(1-\frac{\hat{\Lambda}c^2}{16\pi G\rho_{0}}\right)^3}\left\{1+3ap^{0}\left(\frac{\alpha}{2}-\frac{\beta \frac{\Lambda c^2}{16\pi G\rho_{0}}}{\left(1-\frac{\Lambda c^2}{16\pi G\rho_{0}}\right)}\right)\right\}, \label{m_final}
\end{eqnarray}
where $M_{0}=\frac{3b_0^3}{2R_0^2\left(1+2\frac{b_0^2}{R_0^2}\right)}$ is the mass of the star in the commutative space-time.

Compactness of an object is the ratio of its mass to its radius, $u=\frac{M}{b}$\cite{vishnu,kalam,maurya}. It denotes the gravitational strength of the object and we find from eq.(\ref{b_final}) and eq.(\ref{m_final}),

\begin{equation}
u=\frac{M}{b}=\frac{3b^2}{2 R^2\left(1+\frac{2b^2}{R^2}\right)}-\frac{\hat{\Lambda}b_{0}^2}{6\left(1-\frac{\hat{\Lambda}c^2}{16\pi G\rho_{0}}\right)^2}\left\{1+ap^{0}\left(\alpha-\frac{2\beta \frac{\Lambda c^2}{16\pi G\rho_{0}}}{\left(1-\frac{\Lambda c^2}{16\pi G\rho_{0}}\right)}\right)\right\}
\label{compactness}
\end{equation}
From the compactness factor we can calculate surface redshift, which is also a good indicator of the strength of the gravitational field of the object\cite{kalam,maurya}, i.e.,
\begin{equation}
Z_{redshift}=\frac{1}{\sqrt{1-2 u}}-1  \nonumber \\ 
\end{equation}
We find,
\begin{equation}
Z_{redshift}=\sqrt\frac{F}{D}\left[1+\frac{\frac{ap^0}{2}\frac{\Lambda b_{0}^2}{3}F\left(\frac{\alpha}{2}-\frac{\beta \frac{\Lambda c^2}{16\pi G\rho_{0}}}{\left(1-\frac{\Lambda c^2}{16\pi G\rho_{0}}\right)}\right)}{\left(1-\frac{\Lambda c^2}{16\pi G\rho_{0}}\right)D}\right]-1,
\label{redshift}
\end{equation}
where
\begin{equation} 
F=1+2\frac{b^2}{R^2}~;~~
 D=1-\frac{b^2}{R^2}-\frac{\hat{\Lambda}b_{0}^2}{3}\frac{F}{\left(1-\frac{\hat{\Lambda}c^2}{16\pi G\rho_{0}}\right)} \nonumber
\end{equation}
Note that the compactness factor and the surface redshift(eq.(\ref{compactness}) and eq.(\ref{redshift})) get $a$ dependent corrections. But these correction terms also depend on $\Lambda$. Thus, in the limit $\Lambda \rightarrow 0$, we see that compactness and surface redshift do not get any modifications due to non-commutativity. This is consistent with the result obtained in \cite{neutronstar}.
 
Since the deformation parameter is expected to be of the order of Planck length, we choose $ap^0=0.01$. Let $p^0(=\frac{E}{c})=10^6M_{\odot}$ which is the mass of the black hole, we find $\frac{mc^2}{E}$ to be of the order of $10^{-63}$. Using these, the value of $\alpha$ and $\beta$ (see eq.(\ref{radius_final})) are found to be $0.5$ and $3.5$ respectively. The density of the star at the boundary is taken to be $\rho_b=2\times 10^{17}kg~m^{-3}$\cite{vaidya} in order to find various choices for $\lambda$. The solar mass($M_{\odot}$) is taken to be 1475 meters in natural units.

Mass, radius, compactness and redshift of the neutron star are calculated for various values of $\lambda$ and cosmological constant (see Table\ref{tab2}).
\begin{table}[h!]
\caption{ The mass and radius of neutron star. \label{tab2} }
\centering
\begin{tabular}{|c|c|c|c|c|c|c|}
\hline
\textbf{$\lambda$}&\textbf{$\Lambda(m^{-2})$}&\textbf{$\left(\frac{b}{R}\right)^2$}&\textbf{$b(km)(radius~of~star)$}&\textbf{$\frac{M}{M_{\odot}}$}& compactness & $Z_{redshift}$\\
\hline
\multirow{1}{*} 0.9 & $10^{-10}$ & 0.03362 & 8.664 & 0.27029 & 0.04601 & 0.05233\\\cline{2-7}
    
       &$10^{-11}$& 0.03285 & 8.475 & 0.26502 & 0.04612& 0.04986\\ \cline{2-7}
   
       &$10^{-12}$& 0.03278 & 8.456 & 0.2645 & 0.04613& 0.04962\\ \cline{2-7}
       
       &$10^{-13}$& 0.03277 & 8.454 & 0.2644 & 0.04613& 0.04959\\ \cline{2-7}
       
       &$10^{-14}$& 0.03277 & 8.454 & 0.2644 & 0.04613& 0.04959\\ \cline{2-7}
       
       &$10^{-15}$& 0.03277 & 8.454 & 0.2644 & 0.04613& 0.04959\\ \cline{2-7}
       \hline
\multirow{1}{*} 0.8 & $10^{-10}$ & 0.07419 & 12.118 & 0.7761 & 0.09446 & 0.11709\\\cline{2-7}
       &$10^{-11}$& 0.07252 & 11.869 & 0.7625 & 0.09476& 0.11143\\ \cline{2-7}
   
       &$10^{-12}$& 0.07235 & 11.844 & 0.7612 & 0.09479& 0.11088\\ \cline{2-7}
       
       &$10^{-13}$& 0.07234 & 11.842 & 0.7610 & 0.09479& 0.11083\\ \cline{2-7}
       
       &$10^{-14}$& 0.07234 & 11.842 & 0.7610 & 0.09479& 0.11083\\ \cline{2-7}
       
       &$10^{-15}$& 0.07234 & 11.842 & 0.7610 & 0.09479& 0.11083\\ \cline{2-7}
       \hline
\multirow{1}{*} 0.7 & $10^{-10}$ & 0.12437 & 14.657 & 1.4489 & 0.14581 & 0.20026\\\cline{2-7}
       &$10^{-11}$& 0.12161 & 14.375 & 1.4266 & 0.14638& 0.19026\\ \cline{2-7}
   
       &$10^{-12}$& 0.12134 & 14.348 & 1.42436 & 0.14643& 0.18930\\ \cline{2-7}
       
       &$10^{-13}$& 0.12131 & 14.345 & 1.42414 & 0.14644& 0.18920\\ \cline{2-7}
       
       &$10^{-14}$& 0.12131 & 14.345 & 1.42411 & 0.14644& 0.18919\\ \cline{2-7}
       
       &$10^{-15}$& 0.12131 & 14.344 & 1.42411 & 0.14644& 0.18919\\ \cline{2-7}
       \hline
\multirow{1}{*} 0.6 & $10^{-10}$ & 0.18848 & 16.683 & 2.26984 & 0.20068 & 0.31287\\\cline{2-7}
       &$10^{-11}$& 0.18437 & 16.385 & 2.23955 & 0.20161& 0.29641\\ \cline{2-7}
   
       &$10^{-12}$& 0.18397 & 16.356 & 2.23647 & 0.20169& 0.29483\\ \cline{2-7}
       
       &$10^{-13}$& 0.18393 & 16.353 & 2.23616 & 0.20170& 0.29468\\ \cline{2-7}
       
       &$10^{-14}$& 0.18393 & 16.353 & 2.23613 & 0.20170& 0.29466\\ \cline{2-7}
       
       &$10^{-15}$& 0.18393 & 16.353 & 2.23612 & 0.20170& 0.29466\\ \cline{2-7}
       \hline
\multirow{1}{*} 0.5 & $10^{-10}$ & 0.27409 & 18.341 & 3.2325 & 0.25995 & 0.47801\\\cline{2-7}
       &$10^{-11}$& 0.26823 & 18.039 & 3.19586 & 0.26132& 0.45066\\ \cline{2-7}
   
       &$10^{-12}$& 0.26765 & 18.009 & 3.19212 & 0.26144& 0.44807\\ \cline{2-7}
       
       &$10^{-13}$& 0.26760 & 18.006 & 3.19174 & 0.26146& 0.44781\\ \cline{2-7}
       
       &$10^{-14}$& 0.26759 & 18.005 & 3.1917 & 0.26146& 0.44779\\ \cline{2-7}
       
       &$10^{-15}$& 0.26759 & 18.005 & 3.1917 & 0.26146& 0.44778\\ \cline{2-7}
       \hline
   \end{tabular}
\end{table}
We see that for a fixed $\lambda$, the mass and radius of the star decrease with a decrease in $\Lambda$, whereas the compactness of the star slightly increases with a decrease in $\Lambda$. The surface redshift decreases very slowly with a decrease in $\Lambda$. 

\begin{table}[h!]
\caption{ The maximum mass and radius of neutron star. \label{tab2a} }
\centering
\begin{tabular}{|c|c|c|c|c|c|c|}
\hline
\textbf{$\lambda$}&\textbf{$\Lambda(m^{-2})$}&\textbf{$\left(\frac{b}{R}\right)^2$}&\textbf{$b(km)(radius~of~star)$}&\textbf{$\frac{M_{max}}{M_{\odot}}$}& compactness & $Z_{redshift}$\\
\hline
0.4799 & $10^{-10}$ & 0.29498 & 18.64 & 3.44 & 0.27 & 0.52\\
\hline
0.4565 & $10^{-11}$ & 0.314493 & 18.66 & 3.66 & 0.29 & 0.54\\
\hline
0.45418 & $10^{-12}$ & 0.316506 & 18.66 & 3.68 & 0.29 & 0.54\\
\hline
0.4540 & $10^{-13}$ & 0.316648 & 18.66 & 3.68 & 0.29 & 0.54\\
\hline
0.4540 & $10^{-14}$ & 0.316641 & 18.66 & 3.68 & 0.29 & 0.54\\
\hline
0.4540 & $10^{-15}$ & 0.316641 & 18.66 & 3.68 & 0.29 & 0.54\\
\hline
\end{tabular}
\end{table}

The variation of maximum mass and radius of the neutron star with the cosmological constant is tabulated in Table-\ref{tab2a}. It is seen that as $\Lambda$ decreases the maximum mass and radius of the star increases. Compactness and surface redshift corresponding to the maximum mass and radius of the star are seen to increase as $\Lambda$ changes from $10^{-10}m^{-2}$ to $10^{-11}m^{-2}$ after which they remain constant as $\Lambda$ decreases(see Table-(\ref{tab2a})) which is in accordance with the result obtained in\cite{hendi}. As stated before, since the strong energy condition (see eq.(\ref{P/rho})) is dependent on $R$ which in turn varies with $\lambda$, we do not get a constant upper bound on $\frac{b^2}{R^2}$ for all values of $\Lambda$. Thus, we derive the upper bound for $\frac{b^2}{R^2}$ by imparting the strong energy condition for each value of $\Lambda$.  
The value for $\lambda$ being less than 0.4 violates the causality condition as seen in Table-\ref{tab1}. The above analysis is done for a fixed value of $ap^0=0.01$, where the radius of the star increases as its mass increases.


\begin{figure}[h!]
\centering
\includegraphics[scale=1.2]{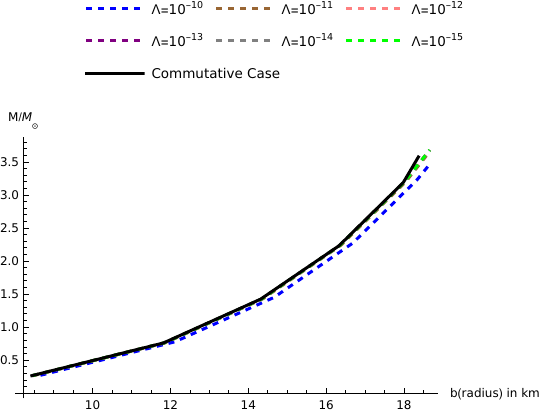}
\caption{$M/M_{\odot}$ vs $b$(radius)}
\end{figure}
In Figure-1, we have plotted the variation of the maximum mass of the star with its radius for $\Lambda$ ranging from $10^{-10}m^{-2}$ to $10^{-15}m^{-2}$ along with the commutative case. The maximum mass is seen to increase with an increase in radius. The plots are very close to each other except for $\lambda=10^{-10}m^{-2}$ where its maximum mass variation with radius is seen to be slightly lower for the non-commutative case than the commutative case.


 Next, we find the variation in the maximum possible mass of the star for a fixed value of $\Lambda=10^{-11}m^{-2}$ by varying the deformation function $ap^0$ (refer Table.(\ref{tab3})). 

\begin{table}[h!]
\caption{Maximum mass of neutron star for different values of $ap^{0}$ .\label{tab3}}
\centering
\begin{tabular}{|c|c|c|c|c|c|c|}
\hline
    \textbf{$ap^{0}$} & \textbf{$\lambda$}& \textbf{$\frac{b^2}{R^2}$}& \textbf{$b(km)$}&\textbf{$\frac{M_{max}}{M_{\odot}}$}& compactness & $Z_{redshift}$\\
\hline
    0.01 & 0.4565 & 0.314493 & 18.662 & 3.65673 & 0.28901 & 0.54366  \\
    \hline
    0.02 & 0.4564 & 0.314583 & 18.709 & 3.6665 & 0.28906 & 0.54385  \\
    \hline
    0.03 & 0.4563 & 0.314674 & 18.756 & 3.67628 & 0.28911 & 0.54404  \\
    \hline
    0.04 & 0.4562 & 0.314765 & 18.803 & 3.68608 & 0.28915 & 0.54424  \\
    \hline
    0.05 & 0.4561 & 0.314857 & 18.849 & 3.69589 & 0.28920 & 0.54444  \\
    \hline
    0.06 & 0.4561 & 0.314835 & 18.895 & 3.7046 & 0.28919 & 0.544395  \\
    \hline
    0.07 & 0.4560 & 0.314929 & 18.942 & 3.71445 & 0.28924 & 0.54459  \\
    \hline
    0.08 & 0.4559 & 0.315024 & 18.989 & 3.7243 & 0.28929 & 0.54479  \\
    \hline
    0.09 & 0.4558 & 0.315119 & 19.036 & 3.73417 & 0.28934 & 0.545002  \\
    \hline
     0.1 & 0.4558 & 0.3151 & 19.081 & 3.7429 & 0.28933 & 0.54497  \\
   \hline
    \end{tabular}
 \end{table}

\begin{figure}[h!] \label{deformation}
\centering
\includegraphics[scale=1.2]{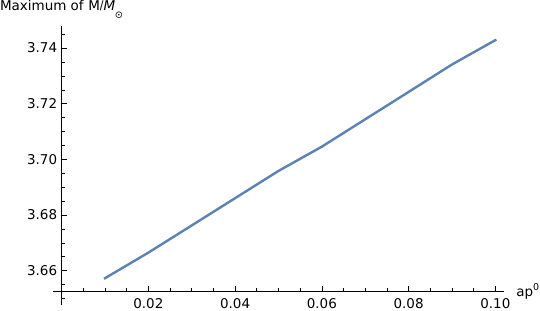}
\caption{$M_{max}/M_{\odot}$ vs $ap^0$(deformation parameter)}
\end{figure}
From Table-\ref{tab3}, we see that the radius and mass of the star increase with an increase in the deformation function $ap^0$. For variation in $ap^{0}$ from $0.01$ to $0.1$, maximum mass increases from $3.65M_{\odot}$ to $3.74M_{\odot}$ and radius changes from $18.6km$ to $19km$.  Compactness factor and surface redshift of the star also increase slowly with an increase in $ap^0$. From the Figure-2, we see a linear change of $\frac{M_{max}}{M_{\odot}}$ with $ap^0$.

\section{Conclusions}

Neutron stars, white dwarfs and black holes produce strong gravitational fields and thus it is natural to study quantum gravity effects in their presence. In this paper, we have investigated the construction and analysis of neutron star in a non-commutative space-time with a non-zero cosmological constant. $\kappa$-deformed space-time, which naturally appears in many quantum gravity models, is the non-commutative space-time in which we construct the TOV equations describing neutron star. After generalizing a metric appropriate for discussing neutron star \cite{vaidya} to $\kappa$-deformed space-time, we have obtained Einstein's field equation in the presence of the cosmological constant. In this approach, one models the non-commutative energy-momentum tensor to be that of a perfect fluid. Solving TOV equations allows us to fix functions appearing in the metric describing the interior of the neutron star in terms of two unknown coefficients. We then use the matching condition between the interior of a neutron star and the exterior region described by $\kappa$-deformed de sitter space-time to fix these coefficients. We further impose strong energy condition and causality condition to find the allowed range of values for the mass, radius, compactness factor and surface redshift of the neutron star. 

Our data shows that as $\Lambda$ decreases from $10^{-10}m^{-2}$ to $10^{-15}m^{-2}$, the mass and radius of the star decreases(see Table-\ref{tab2}). The compactness factor is seen to increase slightly with a decrease in $\Lambda$, whereas the surface redshift decreases with a decrease in $\Lambda$. The maximum mass and radius of the star are found by imposing the strong energy condition everywhere inside the star(see Table-\ref{tab2a}). These quantities are seen to increase as $\Lambda$ varies from$10^{-10}m^{-2}$ to $10^{-11}m^{-2}$ and the remains same with a decrease in $\Lambda$. The maximum mass of the star increases from $3.44M_{\odot}$ to $3.68M_{\odot}$ as the radius changes from $18.63km$ to $18.66km$ for $\Lambda$ varying from $10^{-10}m^{-2}$ to $10^{-15}m^{-2}$. The compactness factor and the surface redshift corresponding to the maximum mass and radius also increases intially and then remain same as $\Lambda$ decreases. Similar results are reported for the study of neutron star with cosmological constant in rainbow gravity \cite{hendi} and in the study of neutron star properties in Einstein-$\Lambda$ gravity without quantum gravity corrections \cite{neutronstarcosmo}. In \cite{hybridneutronstar}, hybrid neutron stars are studied in Einstein-$\Lambda$ gravity where compactness factor and surface redshift are found to decrease with a decrease in 
$\Lambda$ which is in contrast with our result, but variation of maximum mass and radius show similar results as in \cite{hybridneutronstar}. We also see that the non-commutative corrections to the compactness factor and the surface redshift are dependent on the cosmological constant as well. Hence, in the limit $\Lambda \rightarrow 0$, the result reduces to the one obtained in \cite{neutronstar}. The maximum mass versus radius is plotted for fixed values of $\Lambda$(see Figure-1). We see that the maximum mass of the neutron star increases as the radius increases. The result is in accordance with the one reported in \cite{neutronstar} but is in contrast with the result obtained in \cite{neutronstarcosmo,hendi}. The maximum mass versus radius graphs for different choices of $\Lambda$ are seen to be very close to each other(see Figure-1). The behavior of the same for $\Lambda=10^{-10}m^{-2}$ is found to be slightly lower from the commutative case. 

For fixed $\Lambda=10^{-11}m^{-2}$, $ap^{0}$ is varied from $0.01$ to $0.1$ and the maximum mass of the star is seen to increase linearly with $ap^{0}$, which can be seen from Figure-2. The maximum mass increases from $3.656M_{\odot}$ to $3.742M_{\odot}$ as $ap^{0}$ varies from $0.01$ to $0.1$. The radius of the star, compactness factor and its surface redshift are seen to increase with $ap^{0}$. This is in accordance with the result obtained in \cite{neutronstar} where properties of the neutron star are analyzed without cosmological constant in $\kappa$-deformed space-time.

Our study shows the effect on the properties of the neutron star due to the non-commutativity of space-time as well as the cosmological constant. 


\section{Acknowledgement}
BR thanks DST-INSPIRE for support through the INSPIRE fellowship (IF220179). DP thanks  IOE-UOH for support through the PDRF scheme. HS thanks Prime Minister Research Fellowship (PMRF id:3703690) for the financial support. S.K.P thanks UGC, India, for the support through the JRF scheme (id.191620059604). BR, DP, HS and S.K.P  thank E. Harikumar for useful discussions and comments.


\end{document}